\documentclass{biophys-new}
\usepackage[utf8]{inputenc}
\usepackage{graphicx}
\usepackage[colorlinks,allcolors=cyan!70!black]{hyperref}

\usepackage{lipsum}

\title{A hard-sphere model of protein corona formation on spherical and cylindrical nanoparticles}
\runningtitle{Hard-sphere model of NP-protein coronas} 

\author[1,*]{I. Rouse}
\author[1]{V. Lobaskin}
\runningauthor{I. Rouse and V. Lobaskin} 

\affil[1]{School of Physics, University College Dublin, Belfield, Dublin 4, Ireland}

\corrauthor[*]{ian.rouse@ucd.ie}
\papertype{Article}

\begin{document}

\begin{frontmatter}

\begin{abstract}
A nanoparticle (NP) immersed in biological media rapidly forms a corona of adsorbed proteins, which later controls the eventual fate of the particle and the route through which adverse outcomes may occur. The composition and timescale for the formation of this corona are both highly dependent on both the NP and its environment. The deposition of proteins on the surface of the NP is related to processes of random sequential adsorption and, based on this model, we develop a rate-equation treatment for the formation of a corona represented by hard spheres on spherical and cylindrical NPs. We find that the geometry of the NP significantly alters the composition of the corona through a process independent of the rate constants assumed for adsorption and desorption of proteins, with the radius and shape of the NP both influencing the corona. Moreover, we demonstrate that in the condition of strong binding such that the adsorption is effectively irreversible the corona content reflects the protein mobility and concentration in solution rather than their binding affinity.  
\end{abstract}

\begin{sigstatement}
The adsorption of proteins to nanoparticles is key to understanding their biological activity and fate. We present an improved version of the rate-equation based adsorption model that imitates random sequential adsorption and can be applied to curved nanoparticle surfaces. Our results demonstrate that the geometry of the NP alters both the evolution and the final state of the corona when proteins are modelled as rigid particles. In comparison to the previously employed mean-field model, in which proteins are taken to be arbitrarily deformable, we find a significantly longer timespan for the corona to form. We also demonstrate that the protein-nanoparticle interaction can change the corona formation kinetics and the resulting corona content, thus affecting the biological fate of the nanoparticle.
\end{sigstatement}
\end{frontmatter}

\section*{Introduction}

Matter, even elements which are well-known on everyday size scales, takes on new properties on the nanoscale due to increased surface area and quantum mechanical effects, such that even a famously inert metal like gold may become highly reactive in the form of nanoparticles (NP) \cite{phala2007intrinsic}. These new properties, however, potentially lead also to hazards when the NPs are inhaled, absorbed through the skin, ingested, or otherwise find their way into the human body \cite{valsami2015safe,love2012assessing}. A key finding so far has been the observation that a NP in a biological medium rapidly forms a ``corona'' of adsorbed proteins and other molecules \cite{kharazian2016understanding,monopoli2012biomolecular}. This corona is known to be linked to the transport of the NPs around the body, their uptake by cells, and potential adverse outcomes \cite{monopoli2011physical,walkey2014protein,NIENHAUS2020100036}. Thus, the prediction of the content of this corona is key to predicting the safety of novel nanomaterials. 

Qualitatively, the corona may be separated into two classes of bound protein: ``hard'' and ``soft'' \cite{kharazian2016understanding}. The hard corona refers to the proteins strongly bound directly to the surface of the NP, whereas the soft corona is a collection of proteins which are loosely bound on top of the adsorbed protein layer. The content of the hard corona is highly dependent on the material and morphology of the NP, and evolves through the adsorption of proteins upon their diffusion from the bulk, followed by their later desorption and replacement by other proteins over several hours \cite{kharazian2016understanding, monopoli2012biomolecular}. The rate at which proteins desorb is linked to the strength with which they are bound to the NP. The weakly bound proteins are typically smaller and so diffuse faster through the biological medium. Thus, they have a higher rate of collisions with the NP and may feature more in the earlier stages of the corona before being replaced by more strongly-binding proteins which collide less frequently, but do not desorb as rapidly. The phenomenon of protein replacement in the adsorption layer is known as the Vroman effect \cite{slack1995vroman}. The chemical composition of the NP and any modifications to the surface directly influence the strength with which different proteins bind to it, for example, titania NPs are known to be strongly hydrophilic and thus bind more strongly to certain classes of proteins \cite{lundqvist2008nanoparticle}. Intuitively, the curvature of the surface of the NP also influences the strength of the binding: a flat NP will have more of its surface contacting a given protein, and thus experience a stronger attractive force than a curved NP that makes a point contact. We may readily expect therefore that the strength of the binding differs between NPs of the same material but different surface curvatures, i.e., as a function of the geometry of the NP, as is observed in calculations of the binding energy through coarse-grained methods \cite{lopez2015coarse, power2019multiscale}. Beyond this, we expect that the curvature of the surface plays a role in defining how tightly proteins may pack on the surface as a purely geometrical effect \cite{manzi2019simulations,memet2019random}. Indeed, as observed in recent work \cite{madathiparambil2020influence}, spherical and cylindrical NPs of the same material may have have differing corona compositions, which are attributed to a combination of these two effects. 

Given the relevance of the corona to biological outcomes, it is therefore essential to get insight into  the mechanisms of its formation and predict its content. This can be done experimentally through a variety of methods, but these are typically highly sensitive to the exact concentration of proteins used, which may vary significantly between the samples of real biological fluids, and cannot be immediately applied to novel NPs. We therefore turn to theoretical and computational methods to predict both the evolution and the steady state. In principle, this could be achieved by performing a molecular dynamics simulation of the NP immersed in the biological liquid. A brute force solution of the problem is difficult due to the combination of extremely large numbers of atoms and long timescales involved. Atomistic simulations of individual proteins typically require several days to produce a few nanoseconds of simulated time, while the corona may consist of hundreds of proteins and develop over several hours to days \cite{casals2010time}. Thus, to model the actual adsorption-desorption kinetics it is necessary to employ a coarse-grained approach in which the biomolecules are represented by less complex structures that are more amenable to simulation. Even these coarse-grained simulations may be still limited in terms of the amount of computational time taken, and so to probe the dynamics of corona formation on biological timescales one may need to turn to alternate methods, e.g. rate-equation treatments  or different simulation techniques \cite{dell2010modeling, sahneh2013dynamics,shao2016protein,zhdanov2019formation, vilanova2016understanding,vilaseca2013understanding, rabe2011understanding,angioletti14,angioletti18}. 

Previous work on the formation of the corona based on rate equation treatments has primarily focused on a mean-field (MF) approach in which the rate of incoming proteins is reduced by a factor dependent on the total surface area occupied by proteins but further steric factors are not taken into account. That is, if 50\% of the surface area of the NP is occupied by a protein, the binding rate for all incoming proteins is reduced by 50\%. This assumption is justified if the proteins are sufficiently deformable that they pack perfectly tightly without any gaps. If this is not the case, then it is straightforward to envision a situation in which the total uncovered area is sufficiently large to allow the binding of further proteins, but is distributed across a number of gaps each of which are too small individually to admit a further protein. In general, proteins cannot arbitrarily adjust their bound configurations and so it is reasonable to assume that some parts of the NP will be uncovered but unavailable for binding. If we consider in particular globular proteins which may be approximated as spherical, then it is clear that there must be gaps between these proteins due to the impossibility to perfectly pack spheres.  This leads us to consider a model of random sequential adsorption (RSA) with the addition of desorption and surface diffusion, see e.g. Ref\cite{talbot2000car} for an overview of models. In this paradigm,  the proteins are modelled as hard spheres with well-defined positions on the NP. Intuitively, we may expect this hard-sphere (HS) model to produce different results to the MF approach. Indeed, numerical simulations have shown that the HS model predicts effects arising purely from the geometry of the NP and deviations from the mean-field predictions \cite{manzi2019simulations}. Provided that the rate equation method can be adjusted to include these effects, in principle this enables a more physically realistic modelling of the corona formation. However, rate equation treatments have a limitation in that the medium in which the NP is immersed may contain a very large number of different protein species, each of which may bind at an essentially arbitrary orientation relative to the NP with an affinity unique to each orientation. Numerical integration of such a large number of dependent variables rapidly becomes infeasible, while models more complex than the mean-field typically lack analytical solutions for even a single spherically symmetric protein. Thus, we must consider also computational approaches capable of simulating large periods of time, e.g., kinetic Monte Carlo approaches in which the time between the adsorption and desorption of proteins is not directly simulated.

In this work, we employ a combined analytical and computational approach to investigate the effects of the geometry of the NP on the corona composition in terms of how this alters the packing efficiency and the rate at which proteins adsorb to the surface. We demonstrate that the mean-field and hard sphere (HS) models predict significant differences in the timescales and numbers of adsorbed proteins for a simple system of three common blood serum proteins previously employed in studies of corona formation on spherical NPs \cite{dell2010modeling,sahneh2013dynamics}. Using an approach based on scaled particle theory as previously proposed for adsorption onto planar surfaces \cite{talbot1994new}, we develop a rate-equation treatment for the HS model that may be numerically integrated to obtain the evolution of the corona on spherical and cylindrical NPs, finding that the geometry of the NP alters the composition of the corona even when the change in binding affinity due to the surface curvature is neglected. We confirm these results using Kinetic Monte Carlo simulations, finding excellent agreement. Our model highlights the importance of geometric effects, binding strength, and the surface diffusion of the bound proteins.  
\section*{Methods}

\subsection*{Rate equations}

We consider a model of reversible adsorption of proteins to the surface of an NP in which the protein is in physical contact with the surface and occupies an area $A_i$ thus making it inaccessible for other proteins. The surface area of the NP is $A_{\text{NP}}$, and we denote the number of available binding sites for a given protein as $n_i = A_{\text{NP}}/A_i$. We define the number of proteins of type $i$ which are bound to the NP to be $N_i$ and further define the surface coverage of this type to be $s_i = N_i/n_i$. Each protein is further characterised by a rate constant for adsorption, $k_{a,i}$ (in units s$^{-1}$ M$^{-1}$) and desorption $k_{d,i}$ and a bulk molar concentration $[C_i]$. We assume that this bulk concentration remains fixed during the evolution of the corona due to either diffusion of proteins from a reservoir or a sufficiently low concentration of NPs that the effect of binding on the concentration of unbound proteins is negligible.

Once a NP is immersed in a protein solution, the evolution of the resulting corona using a rate-equation treatment is given by
\begin{equation} \label{eq:coronaRatesMF}
\frac{d N_i}{d t} = k_{a,i}  n_{i}  [C_i] P_i(\mathbf{N}) - k_{d,i} N_i ,
\end{equation}
where proteins of type $i$ collide with the NP at a rate given by $k_{a,i} n_i [C_i]$ and $P_i( \mathbf{N})$ is the probability that a colliding protein of this type successfully binds to the surface of the NP as a function of the set of bound proteins, denoted by $\mathbf{N} = (N_1,N_2...N_i)$. We further assume that the protein binds to the surface with unit probability if there is sufficient space for it to do so. Thus, the probability of binding is determined by the probability that a protein randomly positioned on the surface of the NP does not overlap with other proteins.  In the mean-field approach of Refs.\cite{dell2010modeling,sahneh2013dynamics}, this probability is simply equal to the fraction of the surface area of the NP that is not already covered by proteins,
\begin{equation} 
P_i^{MF}(\mathbf{N}) = 1 - \sum_j   \frac{N_j}{n_j }.
\end{equation}
This function is independent of the identity of protein $i$ and thus the rate of binding of all proteins is reduced by the same amount. Using this definition and changing variables from $N_i$ to $s_i$ we obtain
\begin{equation} \label{eq:coronaRatesMFSurface}
\frac{d  s_i}{d t} = k_{a,i}     [C_i] \left (1 - \sum_j s_j \right ) - k_{d,i}   s_i ,
\end{equation}
which is independent of the set of values $n_i$. Thus, in the mean-field model, the corona composition defined in terms of the surface coverages depends only on the concentration of unbound proteins and the rate constants for adsorption and desorption.  The set of linear first-order differential equations given by \eqref{eq:coronaRatesMFSurface} can be solved analytically using standard techniques. Here, we simply give the steady-state corona composition \footnote{Unlike in Ref.\cite{sahneh2013dynamics}, this is the exact solution to the steady-state as we have made the approximation that the concentration of unbound proteins remains fixed.}:
\begin{equation} \label{eq:numBound}
N_i(t=\infty) = n_{i} \frac{ k_{a,j}/k_{d,j} [C_i]}{1 + \sum_j k_{a,j}/k_{d,j} [C_j]} .
\end{equation}   
If the proteins are reasonably strongly binding or have a high concentration then the occupied fraction of the NP surface rapidly approaches unity, that is, the NP becomes completely covered in proteins with no gaps between them. This simple model is valid if the area occupied by a protein can be deformed to enable this optimal packing. In general, however, proteins possess some degree of rigidity preventing this deformation. Consequently, gaps will likely exist between proteins bound to the surface. In the mean field model, the sum of these gaps represents valid area for the binding of further proteins, even if each gap is individually too small to admit a protein. A further limitation of this model is that it assumes that the coverage of the NP reduces the binding rate of each protein identically. Intuitively, we might expect that small proteins are less effected by the increased coverage due to their ability to fit better into gaps between proteins already present on the surface. To quantify these effects, we investigate a model in which the projections of the proteins are taken to be rigid and the proteins are modelled as spherical. We refer to this as the hard-sphere (HS) model, although strictly speaking it is the projections of the proteins which are hard in the sense they cannot overlap or deform. As we are interested only in the monolayer hard corona, the binding process is equivalent to the two-dimensional random sequential adsorption and desorption model which has been thoroughly investigated for planar surfaces. In particular, it has been shown that the acceptance probability for an incoming particle in this model can be approximated by \cite{talbot1994new}
\begin{equation}\label{eq:surfaceCoverageTalbot2}
P_i^{HS}(\mathbf{N}) = \exp(-\mu^{ex}_i(\mathbf{N})/k_B T),   
\end{equation}
where $\mu^{ex}_i$ is the excess chemical potential for the insertion of a new protein of type $i$ as a function of the current state, under the assumption that the state is currently in equilibrium. As before, a protein successfully binds to the NP with probability $P_i^{HS}$ and rejected with probability $1-P_i^{HS}$.  In the case of the adsorption of disks onto a planar surface, it was shown that this assumption is generally quite good even for non-equilibrium states and that the required acceptance probability can be obtained from scaled particle theory (SPT) \cite{talbot1994new}. The probability for a freely-rotating convex particle to have sufficient space to bind is given by
\begin{equation} \label{eq:surfaceCoverageTalbot}
P_i^{HS}(\mathbf{s}) = \left ( 1 - \sum_j   s_j \right ) \exp \left[  -\frac{A_i (\sum_js_j p_j/A_j)^2}{4 \pi (1 - \sum_j s_j)^2}  - \frac{\sum_j s_j( A_i  /A_j + p_i p_j /(A_j  2\pi) ) }{1 - \sum_j s_j}   \right] ,
\end{equation}
where $A_i$ and $p_i$ are the area and perimeter of the projection of protein $i$ onto the surface, $\mathbf{s} = (s_1,s_2...s_i)$. This expression implicitly assumes that the bound particles diffuse on the surface, and we investigate the effects of this assumption later.

\begin{figure}[hbtp]
\centering
\includegraphics[scale=0.65]{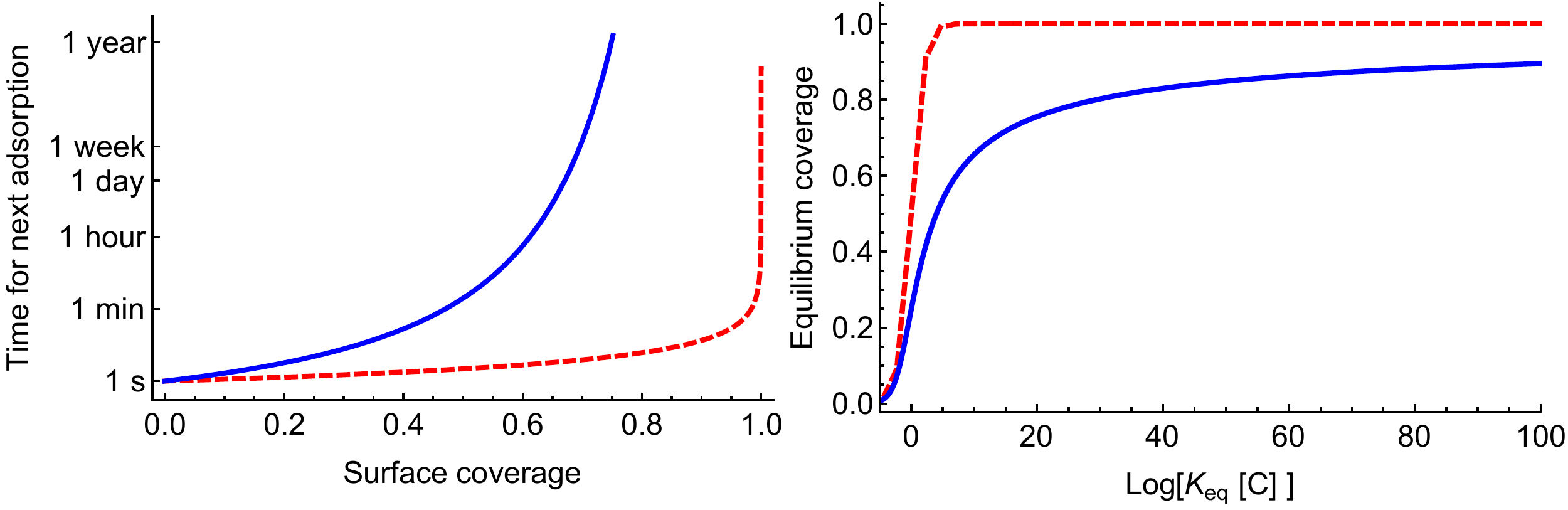}
\caption{Left: The average time taken for a new protein to successfully adsorb to a spherical NP for the hard-sphere (blue) and mean-field (red) models as a function of the occupied area of the NP, for an average adsorption time of 1 second for a bare NP and assuming all proteins are the same size and have a circular binding cross-section. Right: The steady-state surface coverage $s_i$ of a protein with a circular binding cross-section on a spherical NP, defined as a function of the binding equilibrium constant $K_{eq}$ and bulk protein concentration $[C]$. Surface coverages are given for the hard-sphere (blue solid line) and mean-field (red dashed line) models. }
\label{fig:mfhs_coverage}
\end{figure}

In general, the acceptance probability in the HS model is reduced in comparison to that in the MF model, reflecting the fact that an incoming protein is not necessarily able to find a suitably large gap to bind to the NP even if there is sufficient space in total. To illustrate this effect, we consider the special case of a single type of spherically symmetric protein binding to a spherical NP, for which we can assume the projection of the molecule is circular. In this case, Eq.\eqref{eq:surfaceCoverageTalbot} becomes independent of the actual sizes of the protein and NP, but does not reduce to the MF case. In Figure \ref{fig:mfhs_coverage} we plot the average time taken to adsorb a new protein, $1/(k_a [C_i] n_i P_i)$ as a function of the surface coverage for both the mean-field and hard-sphere cases, setting $k_a [C_i] n_i= 1 s^{-1}$ such that one potential adsorption occurs per second. It can immediately be seen that the time taken to adsorb a single new protein rapidly increases even at modest surface coverages in the HS model compared to the mean-field model. Since the formation of the corona likely involves many adsorption events, it follows that even if individual adsorption attempts proceed quicker than the rate of $1 \mathrm{s}^{-1}$ considered in this simple example, the overall formation of the corona will proceed slowly once a moderate surface coverage is reached.  Consequently, the steady-state may not be reached on experimental timescales unless desorption of bound proteins is very fast and the equilibrium surface coverage is low. A rough estimate of the total collision rate $k_{\textrm{coll}} = k_a n_i [C_i]$ for a spherical protein to a spherical NP can be obtained from diffusion theory \cite{chandrasekhar1943stochastic},
\begin{equation}
    k_{\textrm{coll}} \approx 4 \pi (D_{\textrm{NP}} + D_i) ( R_i + R_{\textrm{NP}}) [C_i]
\end{equation}
where $D_j$ is the diffusion coefficient. For a typical protein diffusion coefficient on the order $1 \mu \textrm{m}^2 \textrm{s}^{-1}$ (and neglecting the diffusion of the larger NP), protein concentration $1 \mu \textrm{M}$, with $R_i + R_{\textrm {NP}} = 20$ nm, this implies a few hundred proteins collide with a given NP per second. With this collision rate, we may reasonably expect surface coverages on the order of $0.7$ in the timeframe of a few hours.

It is reasonable to assume that we can employ Eq.\eqref{eq:surfaceCoverageTalbot} to estimate the probability for the successful insertion of proteins into the corona of curved NPs, provided that we correctly calculate $A_i, p_i$ to reflect the projected area of the protein onto the surface of the NP when evaluating the insertion probability. Under this assumption, the evolution of the corona in the hard-sphere model is given by
\begin{equation} \label{eq:coronaRatesHSSurface}
\frac{d  s_i}{d t} = k_{a,i}     [C_i] P_i^{HS} \left (\mathbf{s} \right ) - k_{d,i}   s_i ,
\end{equation}
where $P_i^{HS}$ will in general be a function of both the geometry of the NP and of the proteins. We may therefore expect that, unlike the MF case, the corona composition in this model will depend on the geometry of the NP even if the  parameters  $ka_,k_d,[C_i]$ are kept fixed, since this will alter the values of $A_i,p_i$ used to calculate the value of $P_i^{HS}$. For a single protein, we can calculate the steady-state surface coverage as a function of the equilibrium constant and concentration, as shown in Fig. \ref{fig:mfhs_coverage} and compared to the equivalent prediction for the MF model. It is immediately apparent that this steady-state coverage is much less in the HS model than in the MF, with two key implications. Firstly, we can generally expect fewer proteins to be bound in the HS model, even at very high concentrations or if the binding is particularly strong. Secondly, given an observed surface coverage (e.g. obtained from experiment) and initial concentration of proteins, different equilibrium constants would be obtained depending on which of the two models was employed, as noted in Ref \cite{manzi2019simulations}.

\subsection*{Binding areas}
\begin{figure}[hbtp]
\centering
\includegraphics[scale=0.6]{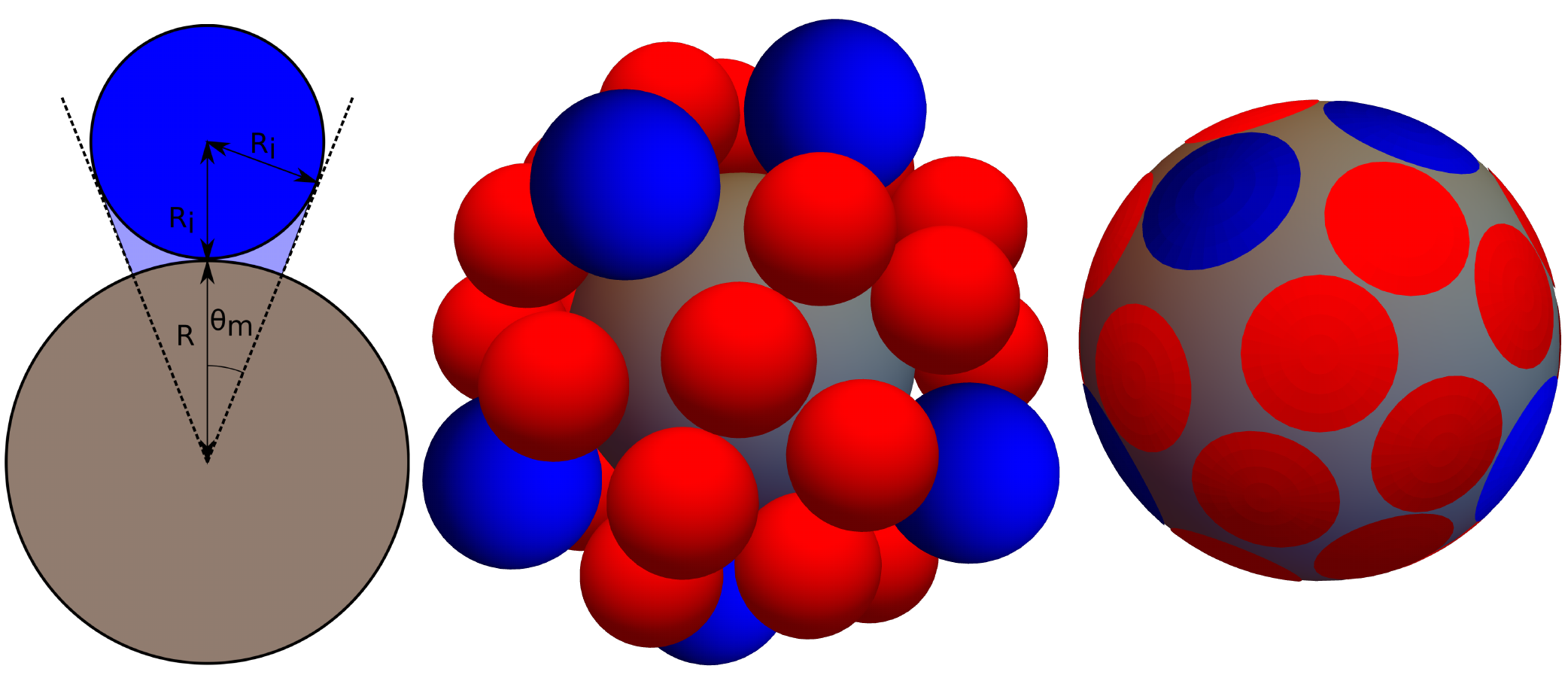}
\caption{ The projection of a spherical protein (blue) of radius $R_i$ onto an NP (gray) of radius $R$. The angle $\theta_m$ is defined as the angle between the line connecting the centres of the NP and protein and the line tangent to the protein and passing through the centre of the NP. The area shaded in pale blue is assumed to be occupied by this protein and unavailable for binding by other proteins, blocking off a region of the surface of the NP. Under this projection, spherical proteins adsorbed to the NP as shown in the centre image can be mapped to two-dimensional regions on the surface of the NP as shown on the right. }
\label{fig:binding_area_def}
\end{figure}

To apply this model, we must define the area $A_i$ occupied by a protein on the surface of an NP. We consider two morphologies of NP: spherical NPs of radius $R_{\text{NP}}$ and rod-like cylindrical NPs of radius $R_{\text{NP}}$ and effectively infinite length. Planar surfaces can be obtained as a limit of either of these with $R_{\text{NP}} \rightarrow \infty$. We assume that all proteins are approximately spherical and characterised by a radius $R_i$, and that a bound protein is in physical contact with the NP. We project the protein onto a curved surface as shown schematically in Fig. \ref{fig:binding_area_def}. For a spherical NP, this occupied area is given in terms of the angle $\theta_m = \sin^{-1} \left(\frac{R_i}{R_i + R_{\text{NP}}}\right)$ by
\begin{equation}
A_i =   R_{\text{NP}}^2 \int^{2\pi}_0 \int_0^{\theta_m} \sin \theta d \theta d\phi,
\end{equation}
such that
\begin{equation}
A_i =  2 \pi R_{\text{NP}}^2 \left [1- \sqrt{1 - \left(\frac{R_i}{R_i + R_{\text{NP}}}\right)^2}  \right ] .
\end{equation}

 For a very large NP, the area tends towards $\pi R_i^2$, i.e., the maximum cross-section of the protein. Conversely, for an NP much smaller than the protein, the number of available binding sites approaches the limiting value of $2$, in agreement with the geometrical constraint that only two spheres may touch at a single point. In all cases, the number of binding sites is given by the surface area of the sphere divided by the area occupied by a protein,  $n_i = 4 \pi R_{\text{NP}}^2 / A_i$. For the HS model we also require the projected perimeter of the protein. By reasons of symmetry, it follows that the projection should be circular and thus the perimeter can be obtained from the area by $p_i = 2 \pi \sqrt{A_j/\pi}$.

\begin{figure}[hbtp]
\centering
\includegraphics[scale=0.5]{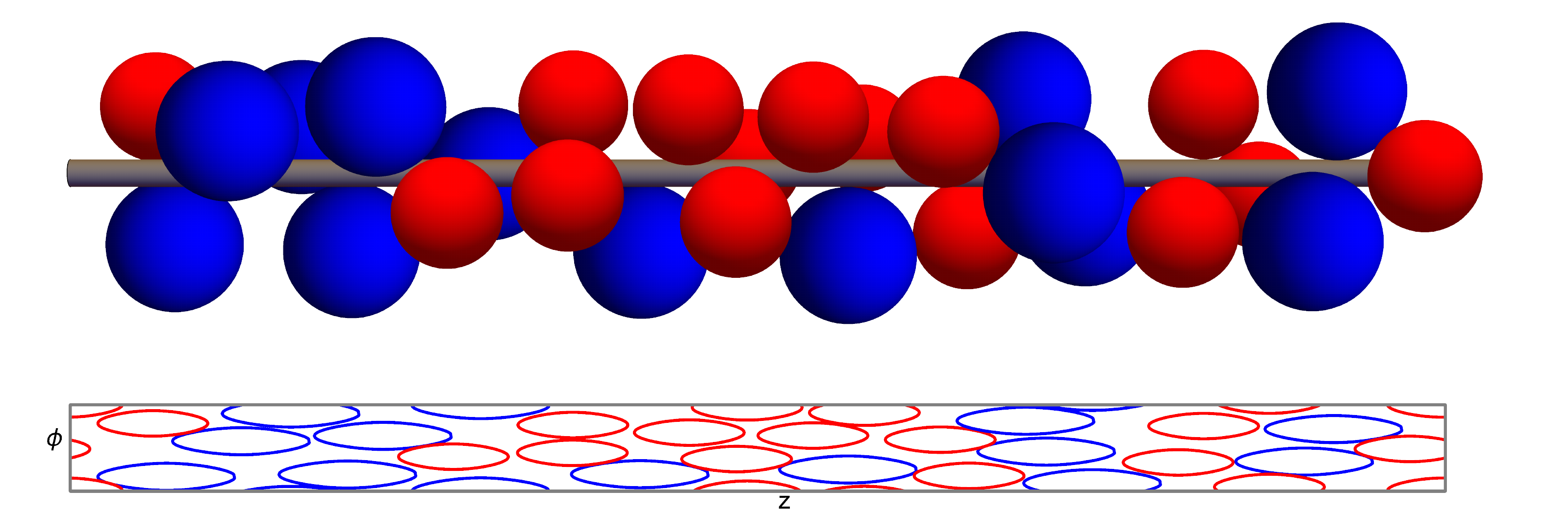}
\caption{A cylinder with multiple adsorbed proteins shown in 3D (top) and the projection of these proteins onto the 2D $\phi z$ plane (bottom).  Positions of the proteins on the cylinder were obtained by kinetic Monte Carlo simulations and their projections in the plane determined following Ref.\cite{memet2019random}, with the periodic boundary conditions shown by included translated protein projections where necessary.}
\label{fig:cylinderprojection}
\end{figure}

Next we turn to a cylindrical NP. The irreversible adsorption of monodisperse spheres onto a cylinder has been recently investigated in Ref.\cite{memet2019random} and demonstrated to be equivalent to the deposition of approximately ellipsoidal shapes onto a planar surface, as depicted in Fig. \ref{fig:cylinderprojection}. The exact area of the projection of the protein onto the cylinder is given by integrating the expression in Ref.\cite{memet2019random} and re-scaling to restore the dependence on $R_{\text{NP}}$,
\begin{equation}
A_i = 4 R_i R_{\text{NP}} \sqrt{\frac{R_{\text{NP}}}{R_i} \left(\frac{R_{\text{NP}}}{R_i}+2\right)} \left(E\left(-\frac{1}{\frac{R_{\text{NP}}^2}{R_i^2}+\frac{2 R_{\text{NP}}}{R_i}}\right)-K\left(-\frac{1}{\frac{R_{\text{NP}}^2}{R_i^2}+\frac{2 R_{\text{NP}}}{R_i}}\right)\right)
\end{equation}
where $E(m)$ and $K(m)$ are the complete elliptic functions of the second and first kind respectively and are given in terms of the parameter $m = k^2$. 

The area $A_i$ can be used to calculate the number of available binding sites for a cylinder of finite (but large) length $L$ and convert the surface fraction into the number of bound proteins according to $n_i = 2 \pi R_{\text{NP}} L/A_i$. To apply the HS model,  we must take into account the fact that the projections are not freely-rotating and thus Eq. \eqref{eq:surfaceCoverageTalbot} does not apply. Following Refs. \cite{martinez2011phase,heyes2018chemical}, the acceptance probability for shapes with a fixed orientation is given by
 \begin{equation} \label{eq:surfaceCoverageEllipse}
P_i^{HS}(\mathbf{N}) = \left(1-\sum _j s_j\right) \exp\left(-\left[\frac{A_i \sum _k \sum _j \frac{s_j s_k A^{ex}_{k,j}}{A_j A_k}}{\left(1-\sum _j s_j\right){}^2}+\frac{\sum _j \frac{s_j A^{ex}_{i,j}}{A_j}+\sum _j \frac{s_j A^{ex}_{j,i}}{A_j}}{1-\sum _j s_j}+\frac{A_i \sum _j \frac{s_j}{A_j}}{1-\sum _j s_j}  \right ] \right ) , 
\end{equation}
where $A^{ex} = \frac{1}{2}(A_{ij} - A_j - A_i )$ is the excess excluded area for protein $j$ due to protein $i$, defined in terms of the co-area $A_{ij}$ which is the area of space in which the centre of protein $j$ cannot be placed without causing an overlap with protein $i$. For the particular case of binding to a cylinder, we approximate this area by the area of an ellipse with axes $R_i + R_j, \tilde{R}_i +\tilde{R}_j$, where
\begin{equation}
\tilde{R}_i = R_{\text{NP}} \sin^{-1} \left( \frac{R_i}{R_i + R_{\text{NP}}} \right),
\end{equation} 
and  $A_{ij} = \pi (R_i+R_j)(\tilde{R}_i+\tilde{R}_j)$.  

It is important to note that, although we have presented the binding areas assuming a spherical protein, the general model proposed here depends only on the projection of the protein onto the surface and not its exact shape. Since Eq.\eqref{eq:surfaceCoverageTalbot} is applicable to freely-rotating convex projections and Eq.\eqref{eq:surfaceCoverageEllipse} to projections of fixed orientation, a wide range of proteins can be modelled using this approach provided that their projected area and perimeter can be calculated. Thus, a complex protein can be treated by estimating binding constants and an associated binding area for each possible orientation relative to the surface of the NP. In this work, we limit ourselves solely to considering spherical proteins to simplify the computational approach described later. We also note that it is the projections of the proteins which cannot overlap, which is a more strict requirement than simply that the proteins themselves do not overlap. This assumption is necessary both to allow small proteins to block larger ones  \cite{talbot1989random} and to ensure that small proteins cannot fit in the gap between larger NPs and the surface of the NP. Physically, this assumption represents that the protein makes contact with the NP over a surface greater than a single point. 
 
\subsection*{Computational methods} \label{section:methods}
The set of equations in the MF approximation can be integrated analytically to produce the evolution of the corona as a function of concentrations and shapes of the proteins. The more complex HS, however, must be integrated numerically. We employ the numerical integration routines in Mathematica 12.1 to do so \cite{Mathematica}. To validate these equations, we also perform simulations of the adsorption process. As discussed previously, even highly coarse-grained models of corona formation using molecular dynamics take an extreme amount of computational time. Thus, we instead employ a kinetic Monte Carlo (KMC) method \cite{voter2007introduction} that can readily be extended to large numbers of proteins. This algorithm samples a set of events -- adsorption or desorption of proteins -- advancing from one to the next without requiring the evaluation of the time inbetween them, allowing for the efficient simulation of a large number of events and thus an extended amount of time. At a given time $t$, the algorithm generates a list of possible events, randomly selects one with a probability weighted by the rate at which that event occurs and the system is updated accordingly. Afterwards, the time is advanced by a random amount given by
\begin{equation}
\Delta t = - \frac{\ln(u)}{k_{tot}} 
\end{equation} 
where $k_{tot}$ is the sum of all event rates and $u$ is a random number uniformly distributed in the interval $[0,1]$. This process iterates until the time reaches a pre-determined stopping point. The possible events correspond to proteins colliding with (and potentially adsorbing to) or desorbing from the surface of the NP. Collisions occur with a rate given by $k_{a,i} n_i [C_i]$ and desorption with a rate $N_i k_{d,i}$. If a desorption event rate is selected, a randomly chosen protein of the specified type is removed from the surface of the NP. Conversely, if a collision occurs, the protein is either accepted or rejected with criteria depending on the model employed. In the MF model, the current occupied surface fraction $\sum_j s_j$ is calculated, and a random number in the interval $[0,1]$ is generated. If this random number is greater than $\sum_j s_j$, the protein is accepted and added to the state of the system. If not, the simulation proceeds to the next event. In the HS model, the location of each protein is explicitly tracked by generating a location for each colliding protein, where these positions are uniformly distributed on the surface of the NP. In the spherical case, a pair of angles $\theta_i,\phi_i$ are randomly generated to produce a uniform distribution of points on the surface of the sphere. This is achieved by drawing two random numbers $u_1, u_2$ from the uniform distribution $[0,1]$ and taking $\phi_i = 2 \pi u_1, \theta_i = \text{cos}^{-1} (2 u_2 - 1)$ \cite{weisstein2002sphere}. A protein of the corresponding type is then inserted at the Cartesian coordinates given by $(R_{\text{NP}} + R_i) [\cos \phi_i \sin \theta_i, \sin \phi_i \sin \theta_i, \cos \theta_i]$ provided that it does not overlap with any existing proteins. An overlap is deemed to have occurred if the condition,
\begin{equation}
\cos^{-1} (\sin \theta_i \sin \theta_j + \cos \theta_i \cos \theta_j \cos(\phi_j - \phi_i)) < \sin^{-1} \frac{R_i}{R_{\text{NP}}+R_i} + \sin^{-1} \frac{R_j}{R_{\text{NP}}+R_j}
\end{equation}
is found to hold for any protein $j$ already adsorbed to the NP and where protein $i$ is the new protein. This ensures that the projections of the proteins onto the surface do not overlap, as assumed in the rate-equation model, and is derived by projection of the centre of the two proteins onto the sphere and calculating the minimum angular distance between them allowable without overlap occuring.

For the cylindrical NP, the position of the protein is defined by the coordinates $\rho_i, \phi_i, z_i$, where $\rho_i = R_{\text{NP}} + R_i$ such that the protein is in contact with the NP and binding occurs only to the curved surface. The angular coordinate $\phi_i$ is chosen from a uniform distribution in $[0,2\pi]$. The final coordinate $z$ is chosen uniformly from the interval $[-L/2,L/2]$, where $L$ is the length of the cylinder. To minimise edge effects, we ensure that $L$ is large compared to the typical radius of the proteins and employ periodic boundary conditions at $\pm L/2$. Unlike in the spherical case, we cannot write down a simple expression to test if a pair of proteins overlap due to the more complex shape of the projection of proteins onto the cylinder. Instead, for a pair of proteins of radii $R_i,R_j$, we determine if overlap occurs by setting the radial co-ordinate for both to $\rho_{i,j} = R_{\text{NP}} + \mathrm{max}(R_i,R_j)$ and determining if this causes a physical overlap between the proteins, i.e. if the distance between their centres is less than $R_i + R_j$. This procedure ensures that a smaller protein cannot bind in the region between a larger protein and the NP, which would be possible if physical overlap was only checked for at their actual location,  is consistent with the definition used for the spherical NP, and avoids a computationally expensive test for overlapping based on the projections of the two proteins onto the surface. By projecting the locations of proteins obtained from simulations onto a planar surface as depicted in Fig. \ref{fig:cylinderprojection} we have determined that this test correctly prevents the overlap of the projected areas of proteins for those considered here, although in general this method may fail if the proteins are much larger than the NP. To implement the periodic boundary conditions with respect to the ends of the cylinders, overlap is tested for the incoming protein against the original set of bound proteins and copies of this set translated by $\pm L$ along the axis of the cylinder. For both spherical and cylindrical NPs, these overlap criteria are expected to produce equivalent results to the theoretical expressions Eq. \eqref{eq:surfaceCoverageTalbot} and Eq. \eqref{eq:surfaceCoverageEllipse} respectively, but the KMC program does not evaluate these pobabilities when assessing if a protein is accepted or not. Thus, this method provides a means to test the validity of these expressions for the systems considered here.

For the majority of the simulations, we assume that once a protein has adsorbed to the NP it remains fixed in place until it desorbs. It is known, however, that the diffusion of proteins on the surface of the NP increases the packing efficiency \cite{ravichandran2000mobility}, and the rate-equation model assumes that the adsorbed proteins are effectively in a fluid phase \cite{talbot1994new}. The desorption and readsorption of proteins produces a restructuring of the surface equivalent to the diffusion, justifying the exclusion of this process. If, however, proteins are sufficiently tightly bound that they do not escape the NP but are still free to diffuse across the surface, then it is of interest to investigate how the surface restructuring impacts the results. To implement this, after each event the proteins are selected in a random order and the angular position of each molecule perturbed by a small random amount. These perturbations are randomly drawn from zero-mean normal distributions with standard deviations given by \cite{apaza2017brownian}
\begin{equation}
    \sigma_\theta = \frac{2 D_s \Delta t}{R_{\text{NP}}^2}
\end{equation}
\begin{equation}
    \sigma_\phi = \frac{2 D_s \Delta t}{R_{\text{NP}}^2} \left(\frac{1}{\theta^2} + \frac{1}{(\theta - \pi)^2}\right)
\end{equation}
where  $\Delta t$ is the time between the most recently applied event and the event before and $D_s$ is the surface diffusion coefficient, which we set to $10^5$  nm$^2 \cdot $s$^{-1}$ for all proteins. A test move is generated and accepted if it does not result in an overlap between proteins and otherwise rejected, with up to four test moves attempted before moving to the next protein, with proteins selected in a random order. This requires re-checking for collisions following each trial perturbation and so dramatically slows the simulation. Thus, we do not employ this procedure for most simulations and have implemented it in the code only for spherical NPs.

\section*{Results}

In this work, we are primarily interested in determining how the geometry of the NP influences the corona, independently of the strength with which a given protein binds to the surface of the NP. Thus, we vary the size and shape of the NP while keeping the concentrations and rate constants for adsorption and desorption for each protein fixed and present results comparing the KMC simulations to the rate-equation model. For consistency with the literature, we primarily employ the same system of proteins as in Ref. \cite{dell2010modeling} using the same rate constants and concentrations quoted there as summarised in Table \ref{tab:proteinData}. This system consists of three proteins: HDL, HSA, and fibrinogen (Fib). Of these, HDL and HSA are globular and so can be reasonably modelled as spherical. Fibrinogen is a rod-like protein, which can be treated using the analytical model but not in the current implementation of the CoronaKMC program for side-on orientations. For the purposes of this work, we simply assume the binding is achieved through an end-on configuration, such that the binding profile is approximately equivalent to that of a globular protein.  The exact identity of these proteins is immaterial as we are focused on the mathematical treatment rather than attempting to match experimental results, and we simply require a range of concentrations and binding constants. Thus, Fib can also be considered to be a large but weakly binding globular protein, with the name used purely to ensure consistency with Ref. \cite{dell2010modeling}. Of the proteins considered here, HDL binds significantly more strongly than the other two, with a characteristic lifetime $1/k_d$ on the order of $\approx 9$ hours compared to $\approx 8$ minutes for HSA and Fib. In certain simulations, we also consider two fictional proteins corresponding to one comparable to Fib but with a higher concentration, and one which is a slightly larger form of HDL. To enable a comparison between NPs of different sizes and morphologies, we present all results in terms of the surface coverages $s_i$, using the relation $s_i = N_i/n_i$ to convert the numbers of bound proteins observed in KMC simulations to the surface coverage. 

\begin{figure}[hbtp]
\centering
\includegraphics[scale=0.6]{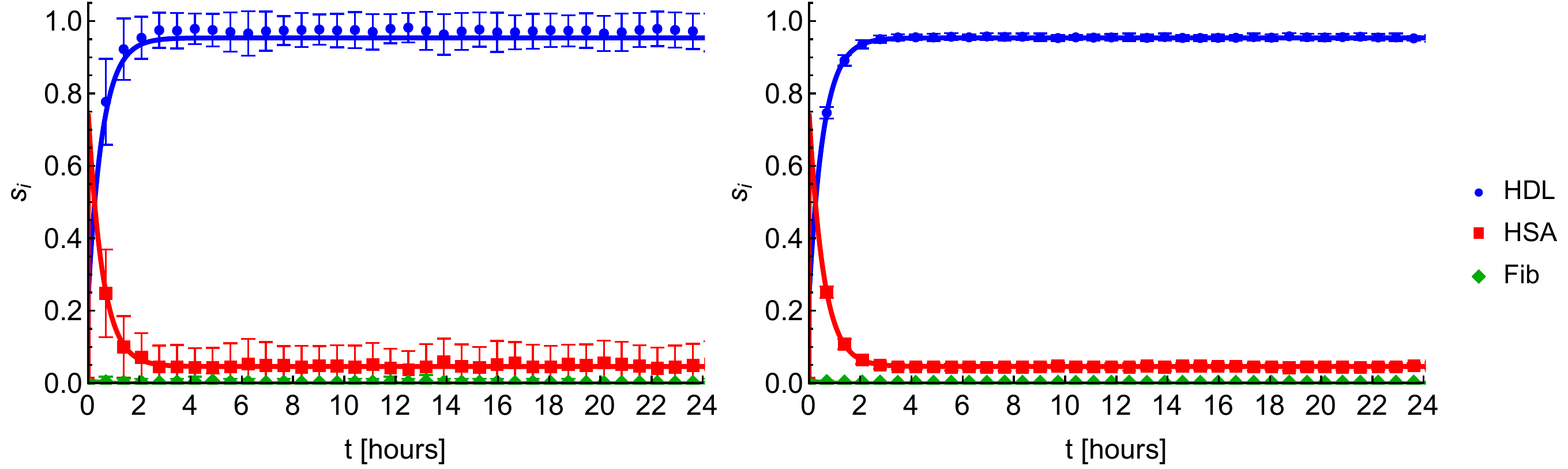}
\caption{ Evolution of the corona in the mean-field model, showing the surface fraction occupied by HDL (blue), HSA (red) and Fib (green). The points show the results obtained from the kinetic Monte Carlo simulations for NPs of radius $5$ (left, 100 iterations) and $50$ (right, 50 iterations) nm, with error bars indicating the standard deviation of these results. The solid lines indicate the analytical prediction obtained by integration of the rate-equation model.}
\label{fig:corona_5and50_nm_mf_kmc}
\end{figure}

First, we confirm that the KMC algorithm can be used to generate results in the mean-field approximation by performing simulations using this algorithm and comparing it to the analytical solution to the rate-equation treatment. We find that the KMC method produces results essentially identical to the analytical method, as can be seen from the clustering of the trajectories around the analytical prediction for spherical NPs of radius both $5$ and $50$ nm in Fig.~\ref{fig:corona_5and50_nm_mf_kmc}. Moreover, it can be seen that the mean surface fraction is independent of the radius of the NP, although the fluctuations around this number are larger for the smaller NP. The coverage for single-particle adsorption on infinite surfaces is known to follow a Poisson distribution \cite{adler2000equilibrium}, for which $\sqrt{ \langle N_i^2 \rangle - \langle N_i\rangle^2  } \propto \sqrt{N_i}$, such that fluctuations in the surface coverage scale as $\sqrt{ s_i/n_i }$.  The true distribution for multi-component adsorption on finite surfaces is of course likely to be much more complex, but heuristically we can expect similar scaling of fluctuations, especially when the steady-state corona consists of essentially only a single protein as in the present case. 

\begin{table}[bt]
\begin{tabular}{l| llll}

    & {[}C{]} [$\mu$M] & $R$ [nm] & $k_a$  [$\times 10^3$ M$^{-1}$ s$^{-1}$]                                             & $k_d$ [    $\times 10^{-3}$ s$^{-1}$] \\
\hline
HDL & 15    & 5    & $30   $ & 0.03   \\
HSA & 600      & 4    & $2.4   $& 2    \\
Fib & 8.8    & 8.3  & $2    $ & 2   \\
FP1 & 600    & 8.3  & $2    $ & 2   \\
FP2 & 15    & 5.5    & $30    $ & 0.03   
\end{tabular}
\caption{Summary of the parameters for the three proteins HDL, HSA and Fib used for the numerical investigation taken from Ref \cite{dell2010modeling}. The proteins FP1 and FP2 are fictional proteins based on modifications of the original three proteins to highlight the effects of varying parameters in the model.}
\label{tab:proteinData}
\end{table}

\begin{figure}[hbtp]
\centering
\includegraphics[scale=0.6]{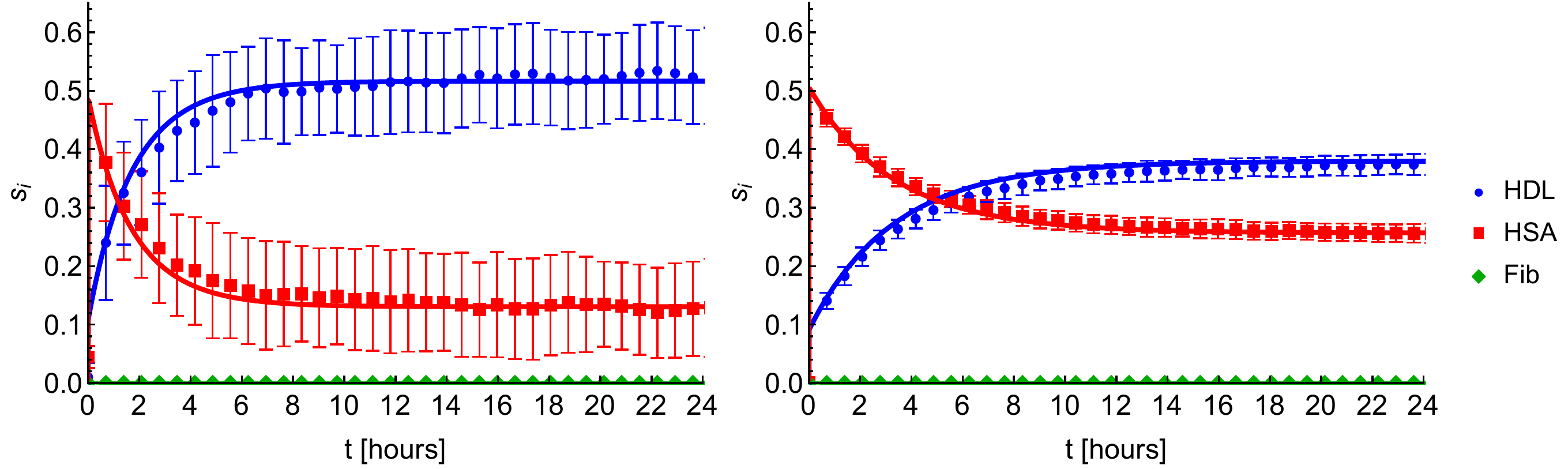}
\caption{Evolution of the corona in the hard sphere model on a spherical NP of radius 5 nm (left) and 50 nm (right). The points show the mean surface fraction  of the adsorbed proteins from the KMC simulations averaged over 100 runs with the error bars indicating the standard deviation. The solid lines indicate the predictions of the rate-equation model.
}
\label{fig:corona_5and50_nm_rsad_kmc}
\end{figure}

\begin{figure}[hbtp]
\centering
\includegraphics[scale=0.6]{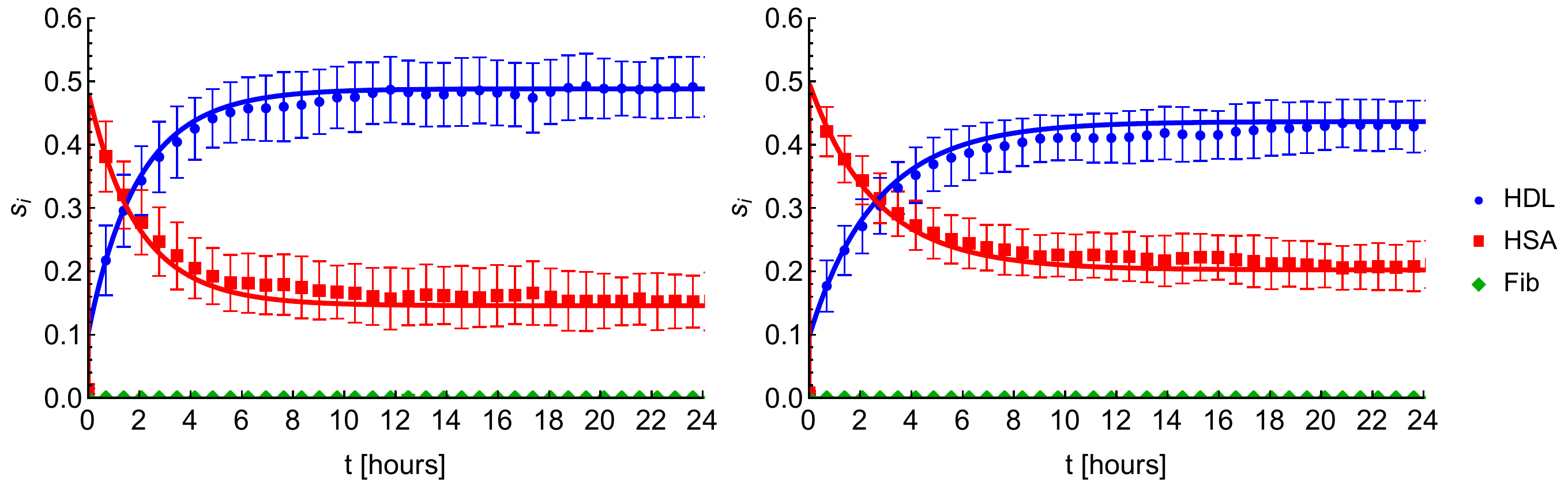}
\caption{Time-dependent surface coverage of proteins on a cylindrical NP of radius 1 (left) and 5 (right) nm in the hard sphere model. The lengths of the cylinders are set to 100 nm with periodic boundary conditions applied. Solid lines indicate the numerical integration of the rate-equation model and points indicate the results of numerical simulations. Error bars indicate $\pm$ one standard deviation as obtained from the simulations.  }
\label{fig:corona_cylinder}
\end{figure}

Next, we perform simulations for this same set of proteins in the HS model. The results for spherical NPs of size $5$ and $50$ nm  are shown in Fig.~\ref{fig:corona_5and50_nm_rsad_kmc}. It is immediately apparent that this significantly alters both the surface coverage and the rate at which the system approaches equilibrium compared to the mean-field model. The analytical model using the surface coverage given by Eq.~\ref{eq:surfaceCoverageTalbot} is in  good agreement with the numerical results. It can be seen that, unlike in the MF model, the surface coverage in the steady-state depends on the radius of the NP. Since we have left the rate constants unchanged, this implies that this is a purely geometric effect arising solely from the different packing efficiencies of spheres on curved surfaces. Likewise, in Fig. \ref{fig:corona_cylinder} we present a comparison between the numerical simulations and the numerically-integrated rate equation treatment for cylinders of radius $R=1$ and $R=5$nm, again finding that the radius of the NP plays a role in defining the evolution of the corona even when rate constants are not altered. In both cases, the time taken to reach the steady-state is on the order of several hours despite the relatively quick desorption of individual HSA molecules, which occurs on the order of minutes. This is most likely caused by the preferential re-adsorption of HSA into the gaps left by its desorption compared to the adsorption of the larger HDL molecules.

\begin{figure}[hbtp]
\centering
\includegraphics[scale=0.6]{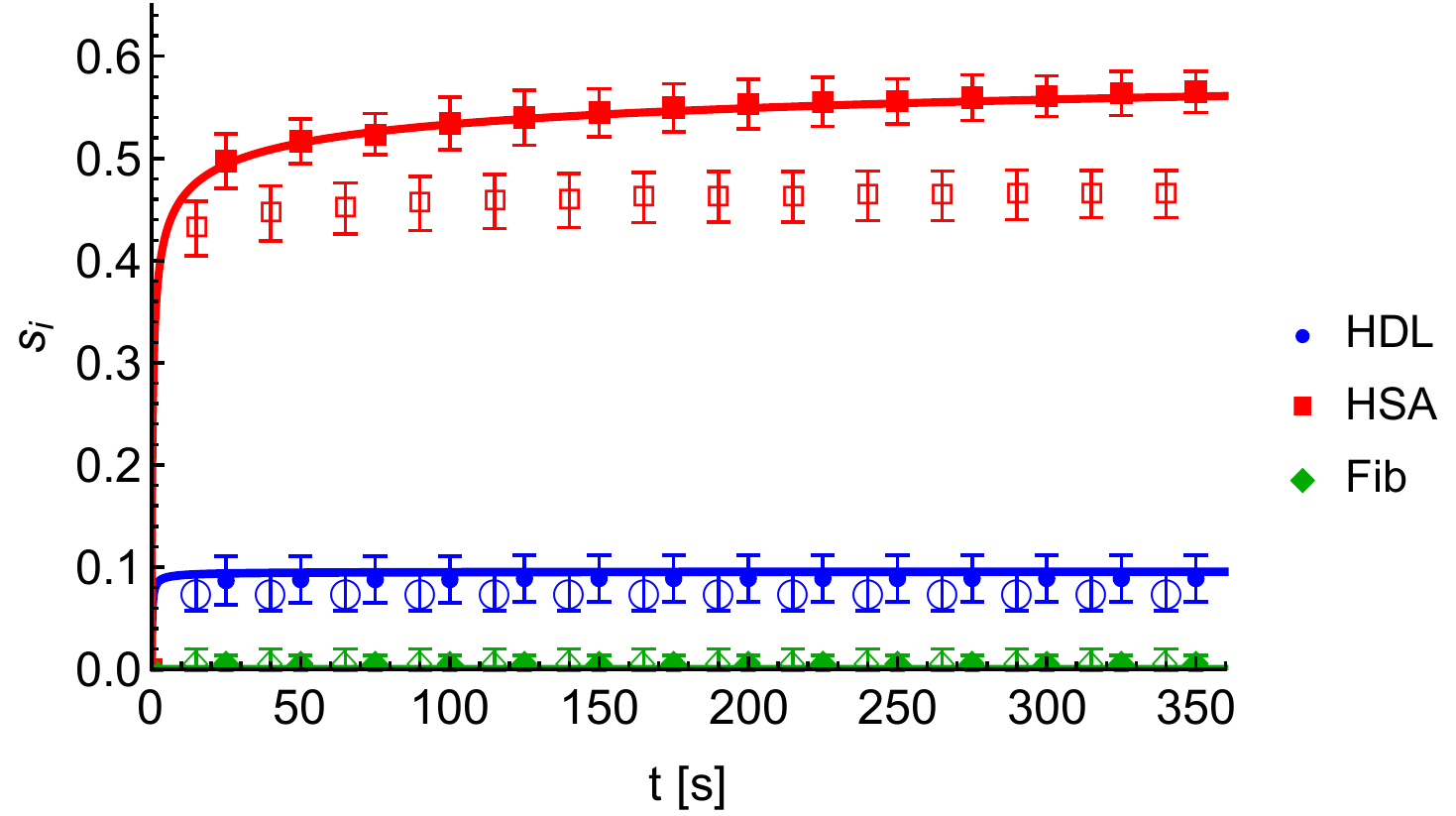}
\caption{Time-dependent surface coverage of proteins on a spherical nanoparticle of radius 20 nm in the hard sphere model in which the adsorption is taken to be irreversible. Solid lines indicate the rate-equation hard-sphere model, solid points indicate the results of simulations with surface diffusion, and open symbols indicate the simulations excluding surface diffusion. 10 iterations of the simulation are performed for each case, with error bars indicating the standard deviation and points the mean.}
\label{fig:corona_irreversible}
\end{figure}

Certain nanomaterials, for example gold, are known to adsorb proteins so strongly that the binding is essentially irreversible, $k_{d,i} \approx 0$ \cite{power2019multiscale}. In this case, we may intuitively expect that the corona primarily represents the proteins which collide with the NP more frequently than others. The rate-equation model proposed previously is in principle still valid if we set $k_d = 0$ for all proteins, however, the probability for successful adsorption of a protein in the hard-sphere model assumes a fluid-like phase of bound proteins \cite{talbot1994new}. If proteins cannot desorb, then the bound layer of proteins does not meet this requirement and so we may expect deviations between the simulations and rate-equation model. The fluid-like behaviour can also be obtained by allowing bound proteins to diffuse on the surface of the NP and so we perform simulations including the effects of surface diffusion. The results in Figure  \ref{fig:corona_irreversible} show a comparison between the simulations with (points) and without (open circles) surface diffusion for the protein set in the absence of desorption. Clearly, the rate-equation model remains accurate when surface diffusion is enabled for irreversible adsorption, but fails to accurately describe the results if the adsorbed proteins cannot diffuse on the surface. Due to the limitations of computational resources, only a small number of simulations were performed for with diffusion, but these confirm the fact that the corona has effectively reached its steady state within $360$ s. Formally, we expect the rate-equation model in the limit of non-desorbing proteins to produce inaccurate results after an extreme period of time, as it can straightforwardly be seen that a steady-state solution with $k_{d,i} = 0$ can only exist when the surface of the NP is completely covered in proteins. Geometrically, this optimum packing cannot be achieved for spherical proteins. In practice, however, we find that for the set of proteins considered here the numerical integration of the rate-equation treatment provides physically realistic results even for timespans of up to one year, far exceeding experimental timescales. 

\begin{figure}[hbtp]
\centering
\includegraphics[scale=0.6]{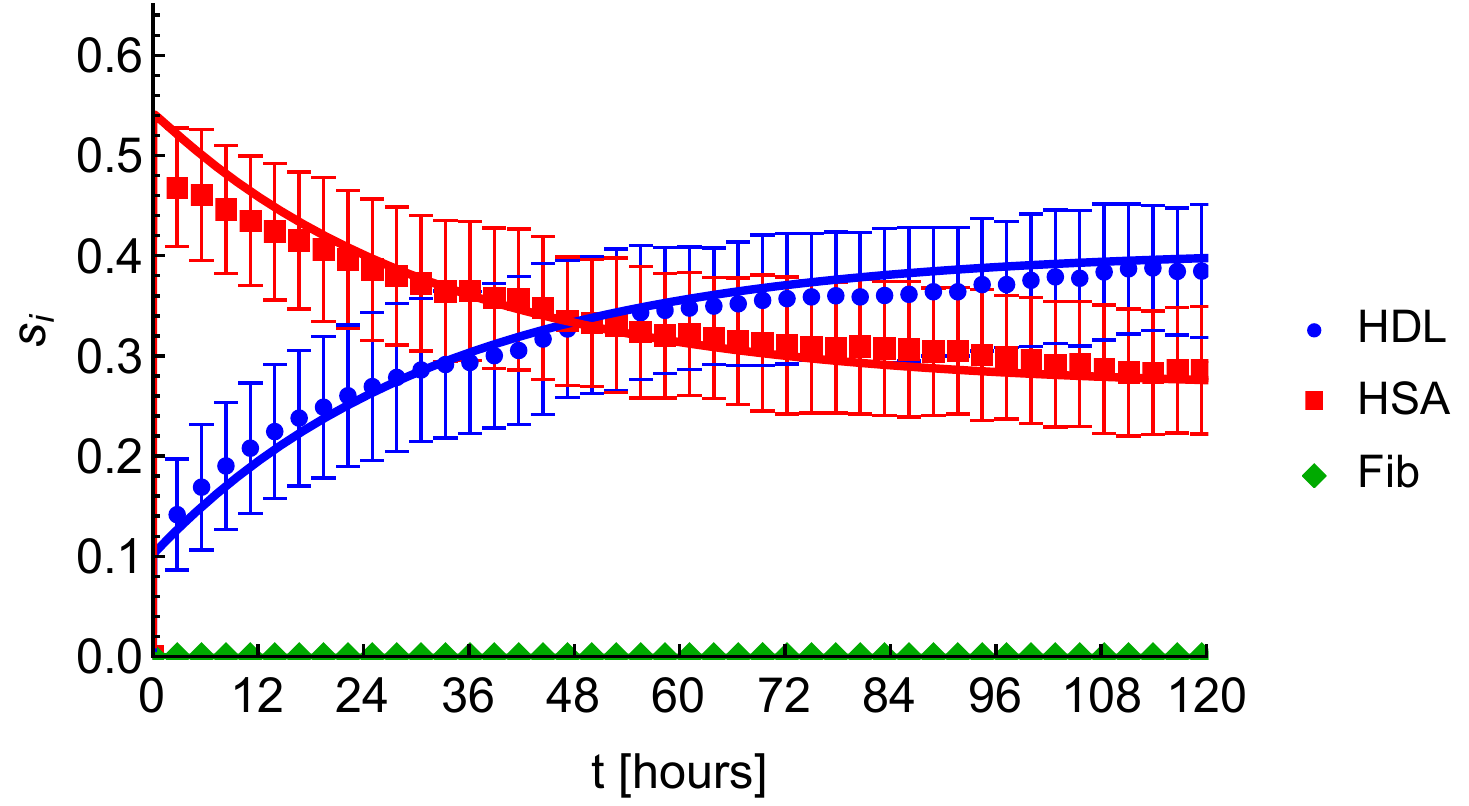}
\caption{Time-dependent surface coverage of proteins on a spherical nanoparticle of radius 10 nm in the hard sphere model, with desorption rates decreased by a factor of 10 compared to the values presented in Table \ref{tab:proteinData}. Points and error bars show the mean and standard deviation obtained from 100 iterations of the kinetic Monte Carlo simulations, lines give the result of numerical integration of the rate equations. }
\label{fig:corona_slowdesorb}
\end{figure}

For small but non-zero desorption rates, we find again that the evolution of the corona is significantly delayed. Even a relatively minor decrease in the desorption rates by a factor of 10 (Figure \ref{fig:corona_slowdesorb}) results in the steady-state corona taking several days to form in the hard-sphere model. A slightly worse agreement between the KMC and rate-equation models can be observed in the initial stages of Fig. \ref{fig:corona_slowdesorb}. Taking into account the results found for irreversible diffusion, it is likely that this is due to the inability of the layer of bound proteins to restructure sufficiently quickly during the initial adsorption to provide the fluid-like properties necessary for the model to hold. We have performed further simulations of the initial stage of formation including diffusion and find a better agreement when this is included.

\begin{figure}[hbtp]
\centering
\includegraphics[scale=0.6]{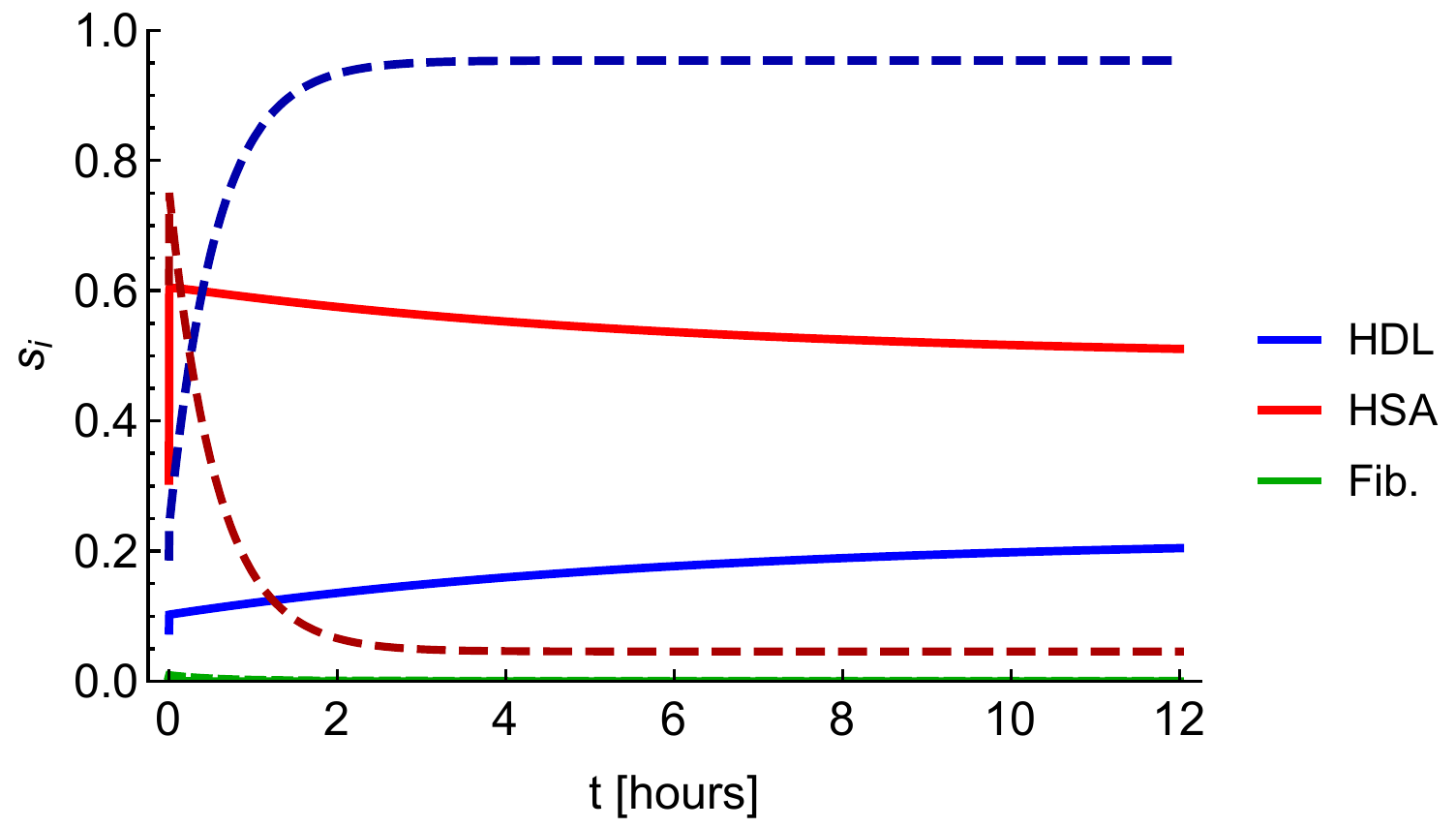}
\caption{Time-dependent surface coverage of proteins on a spherical nanoparticle of radius 10 nm in the hard sphere model (solid lines) and mean field model (dashed lines and darker shades) obtained from the numerical integration of the rate equations. Adsorption rates for all proteins are increased by a factor of 1000 compared to the values presented in Table \ref{tab:proteinData}. }
\label{fig:corona_fastadsorb}
\end{figure}

The adsorption rate constants listed in Table \ref{tab:proteinData} are quite low, corresponding to (on average) less than one potential adsorption event per second for Fib interacting with an NP of radius 10 nm, and even the highly abundant HSA undergoes less than one hundred events per second. To demonstrate the effect of either significantly increasing the concentration or mobility of proteins, we next investigate the outcome of increasing  all adsorption rate constants $k_a$ by a factor of $1000$. Although this produces unrealistically high collision rates, it serves to demonstrate an important point. The results of numerically integrating the rate equations for both the HS and MF models are shown in Fig. \ref{fig:corona_fastadsorb}. \footnote{Due to the high rate of adsorption, simulating the corona using the KMC method for the timespans required to reach steady-state is impractical for this system even with diffusion disabled.} In the MF model, there is essentially no effect on the evolution of the corona compared to the results obtained using the original rate constants. In contrast, however, we observe a significant difference in the HS model: the steady state corona is now primarily composed of the more abundant HSA rather than the more strongly binding HDL. Furthermore, the corona still takes a significantly long time to reach this steady state, with equilibrium not reached after 12 hours, although only minor changes take place after this point.

\begin{figure}[hbtp]
\centering
\includegraphics[scale=0.6]{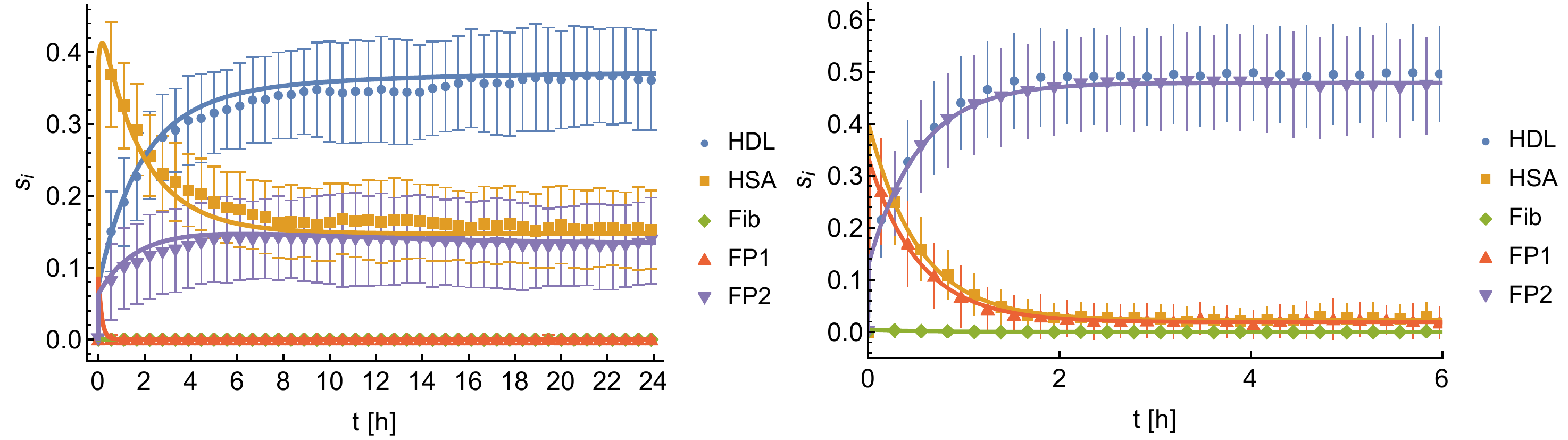}
\caption{Time-dependent surface coverage of proteins on a spherical nanoparticle of radius 10 nm in the hard sphere model (left) and mean field model (right), including additional proteins: FP1, a variant of Fib with a much greater concentration, and FP2, a variant of HDL with the radius increased by 0.5 nm. Lines show the results of numerical integration of the rate equation treatment and points show mean results from 100 iterations of numerical simulations, with error bars indicating the standard deviation.}
\label{fig:corona_fictionalproteins}
\end{figure}

As a further test of the model, we add two additional proteins to the set originally considered and re-calculate the expected corona composition. The first of these two proteins, FP1, is simply the original set of parameters for Fib with the concentration increased to be equal to that of HSA, while the second additional protein FP2 is equivalent to HDL but with a slightly greater radius $R_i = 5.5$ nm in comparison to the original radius of $5$ nm. Results are plotted in Fig. \ref{fig:corona_fictionalproteins} for both HS and MF models. In the HS model, it can be seen that even significantly increasing the concentration of Fib is not sufficient to cause it to be expressed in the long-term corona, as although it is initially adsorbed it is rapidly replaced by more strongly adsorbing proteins (HDL, FP2) or those with similar adsorption characteristics but a smaller size (HSA). Indeed, although both HSA and FP1 have nearly identical values of $k_a [C]$ and $k_d$, it is notable that of the two only HSA is exhibited in the steady-state corona. Meanwhile, FP2 has adsorption kinetics quite similar to that of the original protein HDL, but due to its increased size is less strongly expressed in the corona. Conversely, in the mean-field model, HDL and FP2 have identical surface coverages as a result of the fact in this model these coverages are independent of the radius of the protein. Thus, the difference in surface coverage observed in the HS model is due to the radius-dependent insertion probabilities. Likewise, in the MF treatment HSA and FP1 exhibit very similar profiles, in contrast to the size effects observed in the HS model.

\section*{Discussion}

From the results shown in the previous section, it is clear that the timescale for the formation of the corona varies significantly with the model used to describe how the rate of binding is reduced by pre-existing proteins. At one extreme, if proteins are allowed to move freely around the surface and deform to an effectively arbitrary degree as in the mean field model, the corona approaches equilibrium rapidly. At the other extreme in which the proteins are rigid and fixed in place, the time taken for the corona to reach equilibrium is dramatically increased for the set of proteins considered here. In both cases, the results of the KMC simulations can be adequately explained using analytical models of the adsorption process. Strictly speaking, the model of Ref. \cite{talbot1994new} used for the HS approach is valid for an equilibrium state, but appears sufficiently accurate for the dynamical states observed here. Qualitatively, both HS and MF models predict reasonably similar behaviour for the evolution of the corona -- a sharp initial rise in HSA followed by slow replacement by HDL until the steady-state is reached, although on significantly different timescales. With the set of rate constants used here, the NP corona in the HS model takes several hours to reach equilibrium as a consequence of the decreased rate of adsorption at higher coverages. The difference in timescale from the HS approach arises purely from the decreased probability for proteins to find sufficient space on the NP to adsorb, and is independent of the rate constants. Indeed, even increasing the adsorption rate by a factor of one thousand still results in the HS model predicting timescales for corona formation on the order of twelve hours. 

In contrast to the MF model, the HS model predicts that the size of the NP relative to the size of the proteins significantly alters the evolution of the corona. This is especially apparent in the results shown in Fig. \ref{fig:corona_fictionalproteins}, where simply increasing the radius of a protein by $0.5$ nm significantly reduces the extent to which it features in the corona.  Furthermore, it can be seen that in the MF model the more strongly-binding protein essentially entirely excludes a weaker-binding one,  whereas in the HS model a weakly-binding but small protein may co-exist in the steady-state corona with a larger, more strongly binding protein.  The question therefore remains as to which model -- MF or HS --  is more appropriate to describe the formation of protein coronas under biological conditions. In reality, proteins are neither completely deformable as assumed for the mean-field approach, nor completely rigid as in the hard-sphere model, and so we may expect that the actual timescales are not as slow as predicted in the HS model.  

A key limitation of the analytical approach -- whether mean-field or hard-sphere -- is the fact that obtaining the solution to the set of differential equations rapidly becomes difficult as the number of proteins considered increases. For the results shown here, we have included a set of only three to five proteins, for which the equations may be numerically integrated using e.g. Mathematica. For the mean-field model at fixed concentrations of unbound proteins as considered here, an analytical time-dependent solution can be straightforwardly obtained using matrix methods, producing solutions which are sums of exponential functions. However, both the numerical and analytical approaches to solving these equations rapidly become more complex as the number of proteins grows and if we must also consider multiple orientations for each protein then even this simplified model may stretch the limit of computational resources, especially if the numerical integration must cover a large timespan. The KMC simulations, in contrast, scale reasonably well with respect to the number of proteins but become increasingly slow for larger NPs when a significant number of proteins are bound, especially if surface diffusion is enabled.

We must also address the limitations of the current model. We have assumed for the HS model that the proteins can be represented by hard disks with diameters calculated from that of the corresponding hard sphere. It has previously been shown in the RSA model that a true hard-sphere model exhibits different dynamics as a result of the decreased ability of smaller proteins to block larger ones, and, conversely, the ability of smaller proteins to bind in gaps between the NP and larger proteins \cite{talbot1989random}. However, that work suggests that for proteins a hard disk model is more reasonable, and it is unlikely that proteins make only a single point contact with the NP as the true hard sphere model would require. For the system of proteins considered here, the sizes are sufficiently close that there is not likely to be a significant difference caused by the differences between these two models. A further complication is that protein shapes may deviate significantly from spherical and deform to improve their binding ability, with alternate binding profiles shown to result in increased coverage \cite{maffre2014effects}. While not accounted for in the present model, we note that the surface function (Eq.~\ref{eq:surfaceCoverageTalbot}) derived from scaled particle theory is in principle valid for an arbitrary convex binding area. It is therefore likely that more complex protein shapes can be modelled by finding the area and perimeter of the projection of the protein onto the surface of the NP, treating different orientations as being effectively different proteins. This would further allow for a differentiation between different binding profiles, e.g. head-on vs. side-on for ellipsoidal proteins. Here, we do not investigate this further, but the CoronaKMC tool provided in the UnitedAtom package supports proteins with multiple binding profiles by representing these as hard-spheres of varying size.

The  concentration of unbound proteins in this work and the rate at which they diffuse to the NP surface is here assumed to be fixed. Previous work in the MF model has indicated that finite rates of diffusion to the surface of the NP influence the dynamics of the corona formation \cite{zhdanov2016kinetics}, which could be represented here by a decrease in the adsorption rate $k_a$. Likewise, it is reasonable to assume that the dynamics will be altered if the concentration of NPs $[NP]$ is sufficiently high that substantial protein depletion in the environment occurs, i.e., if $[NP] n_i$ is of the same order of magnitude as the concentration of the protein.  This latter effect can be be accounted for in the rate equation model by replacing $[C_i] \rightarrow [C_i](0) - [NP] n_i s_i(t)$, and in the KMC simulations by dynamically updating the concentrations of proteins based on the number currently bound to the NP. A further possible extension would be to include effects such as three-body interactions in the KMC simulations to produce better agreement with experiment as in Ref \cite{vilanova2016understanding}. Our model also does not capture more complex interactions between the proteins and their environment as considered in the dynamic density functional theory of Angioletti et al \cite{angioletti14,angioletti18}, which indicates that for an accurate model of corona formation it may be necessary to compute more physically realistic models of the protein density surrounding an NP, especially for NPs surrounded by a permeable gel. In the present case, however, we operate under the assumption that the NP and proteins in solution are sufficiently mobile that no significant concentration gradients develop.

It is important to note that the HS model, in which the proteins are represented as hard spheres, is conceptually quite similar to mesoscopic molecular dynamics simulations of the formation of protein coronas as in \cite{tavanti2015competitive,vilanova2016understanding,vilaseca2013understanding,lee2020effects}. In future work, it would be interesting to compare the predictions of this model to such simulations. However, the timescales predicted for the system to reach equilibrium are significantly longer compared to the timescales for individual events. Here, we assumed that proteins collide with the NP at a typical rate on the order of 1/second, and equilibrium takes $\approx 20000$ seconds to reach. To get more realistic estimates of the corona formation times, the collision rates from real protein and NP concentrations should be used. On the other hand, it may therefore be possible to simulate only the initial formation of the corona using such simulations and require the HS model to extrapolate to greater timespans. A similar approach has been employed in Ref \cite{vilanova2016understanding}, where the MF model has been fitted to the results of the simulation to extrapolate to greater time periods than can be achieved through MD simulation. Based on the present work, in which the MF and HS models predict both very different timescales for formation based on the same rate constants and concentrations and differing steady-state corona compositions, we conclude that extraction of rate constants by fitting to numerical data (obtained through experiment or simulation) must be done with care to ensure that the correct underlying model is chosen. That is, attempting to fit the MF model to results obtained in the HS model would result in significantly different rate constants to the underlying values, and vice versa. Thus, it is vital that experimental results are interpreted by applying the correct model to obtain physically meaningful rate constants.

\section*{Conclusion}

 Based on the random sequential adsorption and desorption paradigm,  we have developed a rate equation model for the evolution of the nanoparticle-protein corona for proteins modelled as hard spheres  binding to to spherical and cylindrical NPs. We have demonstrated that the geometry of the NP directly influences the composition of the corona through altering the efficiency with which proteins can pack and new proteins can adsorb to the NP, and that this results in geometry-dependent corona compositions even when the rates at which proteins adsorb or desorb from surface sites are kept fixed. We have found that these effects can be explained using scaled-particle theory for both spherical and cylindrical NPs as long as there is either desorption or diffusion of adsorbed proteins to ensure a fluid-like surface layer.  Our results pave the way for an improved understanding of experimental results and the ab initio prediction of protein corona compositions. 

\section*{Author Contributions}

IR and VL designed the research. IR carried out all simulations, analyzed the data. IR and VL wrote the article.

\section*{Acknowledgments}

We are grateful for funding from EU H2020 grants SmartNanoTox (grant agreement No. 686098) and NanoSolveIT (grant agreement No. 814572).

\bibliography{refs}

\end{document}